\documentclass[prb,reprint,superscriptaddress]{revtex4-1}
\usepackage{amsmath}
\usepackage{graphicx}
\usepackage{epstopdf}
\usepackage{color}

\begin{document}

\title{Phase diagram and edge states of the $\nu=5/2$ fractional quantum Hall state with Landau level mixing and finite well thickness}

\author{Anthony \surname{Tylan-Tyler}}
\affiliation{Department of Physics and Astronomy, Purdue University, West Lafayette, Indiana 47907, USA}

\author{Yuli \surname{Lyanda-Geller}}
\affiliation{Department of Physics and Astronomy, Purdue University, West Lafayette, Indiana 47907, USA}
\affiliation{Birck Nanotechnology Center, Purdue University, West Lafayette, Indiana 47907, USA}

\begin{abstract}
The $\nu=5/2$ fractional quantum Hall effect is a system of intense experimental and theoretical interest as its ground state may host non-abelian excitations, but the exact nature of the ground state is still undetermined. We present the results of an exact diagonalization study of an electron system in the disk configuration including the effects of Landau level (LL) mixing and the finite thickness of the quantum well confining the electrons. The degeneracy between the two leading candidates for the ground state,  the Pfaffian and anti-Pfaffian, is broken  by interactions with a neutralizing background, in addition to the inclusion of two- and three-body interactions via LL mixing. As a result of the neutralizing background in the disc configuration, there is a phase transition from the anti-Pfaffian to the Pfaffian as LL mixing is turned on, in stark contrast to what is observed in a spherical geometry.  The LL mixing leads to an increased charge $e/4$ quasihole size. LL mixing interactions are also shown to overcome the effects of edge reconstruction.  Due to finite thickness effects, these properties are enhanced dramatically. We also find that only the Pfaffian and anti-Pfaffian states continue to possess energy gaps at finite width, while gaps for compressible stripe states close.
\end{abstract}

\maketitle

The $\nu=5/2$ fractional quantum Hall effect (FQHE) \cite{5/2Observation1,5/2Observation2} is the only observed FQHE   with an even denominator.  Thus, it falls outside of the Laughlin \cite{Laughlin} and standard composite fermion \cite{Jain} pictures which describe the odd denominator states.  This has resulted in exploring composite fermion pairing effects \cite{CooperPairing}. The leading candidate for the ground state is then the Moore-Read Pfaffian state \cite{MooreRead,StatPairing1,StatPairing2}.  As this state is not particle-hole symmetric, its particle-hole conjugate, the anti-Pfaffian \cite{aPf1,aPf2}, must also be considered as a candidate ground state of the system. These states are of experimental and theoretical interest as their excitations obey non-abelian anyon statistics \cite{MooreRead,nonabelian1,nonabelian2,nonabelian3}. The specific properties of the FQHE system, such as the structure of the edge states,  are determined by which of the two states is realized. However, in the absence of particle-hole symmetry breaking, the Pfaffian and anti-Pfaffian are degenerate. Since it is well
known \cite{StatPairing1,StatPairing2} that the Moore-Read state is the exact zero-energy solution to a certain repulsive three-body Hamiltonian, and the anti-Pfaffian is the solution to its
particle-hole conjugate, there has been a great interest in calculating diagrammatic expansion terms of the Coulomb interaction which include three-body interactions with virtual excitations to other Landau levels (LL) \cite{Pseudo1,Pseudo2,Pseudo3,Pseudo4,Pseudo5} and many numerical calculations attempting to determine the ground state \cite{ED1,ED2,ED3,DMRG,ED4}.  The effects of the finitequantum well thickness  is believed to have stabilizing effects on the $\nu=5/2$ FQHE and is also a subject of intense investigations \cite{Pseudo3,ED4,FiniteThickness1,FiniteThickness2,FiniteThickness3}.

In order to improve the understanding of these experimental systems, we perform an exact diagonalization study of an electron gas (EG)  in a perpendicular magnetic field a distance $d$ above a disk of neutralizing charge \cite{1/3Disk,5/2Disk}. The 2D gas is hosted by an infinite square well potential of width $w$.  Our choice of the disk configuration allows us to utilize the diagrammatic expansion of the Coulomb interaction for LL mixing \cite{Pseudo3} in a system mimicking realistic experiment.  From our simulations, we extract three primary findings.  First, there is a phase transition from the anti-Pfaffian to the Pfaffian as LL mixing is turned on as a result of the neutralizing background, in stark contrast with what is observed in the spherical geometry \cite{ED4}.  Second, motivated by analytic work \cite{LLMixingHoles}, we find that there is an increase in the size of the charge $e/4$ quasiholes as LL mixing is turned on.  Third, we observe that the LL mixing improves the signature of the edge states as well as overcoming the edge reconstruction shown in Ref. \onlinecite{5/2EdgeReconstruction} in the absence of LL mixing.  The finite thickness also acts to improve the features of the incompressible states when compared to a purely 2D case. In particular, only the MR and anti-Pfaffian states continue to possess energy gaps at finite width, while gaps for the compressible stripe states close.

The Hamiltonian which describes our system is given by
\begin{eqnarray}
\label{hamiltonian}
\hat{H}&=&\sum_{i}U_i\left(d\right)a^\dagger_ia_i+\sum_{i,j,k}V^k_{ij}a^\dagger_{i+k}a^\dagger_{j-k}a_ja_i\nonumber\\&+&\kappa\sum_{m<n,M}V_M^{(2)}P^M_{\left(m,n\right)}+\kappa\sum_{m<n<o,M}V_M^{(3)}P^M_{\left(m,n,o\right)}.
\end{eqnarray}
The operators $a_i^{(\dagger)}$ are the creation (annihilation) operators for a single particle state with angular momentum $i$.  The terms $U_i\left(d\right)$ \cite{1/3Disk,5/2Disk} are the interactions between electrons and the neutralizing background, and $V_{ij}^k$ are the matrix elements of the Coulomb interaction. The terms linear in the LL mixing strength $\kappa=me^2\ell_B/\epsilon\hbar^2$ are then the Haldane pseudopotentials \cite{Haldane} describing the diagrammatic expansion of the Coulomb interaction, specifically of the form used in Ref. \onlinecite{Pseudo3} which includes finite thickness effects. The $V_M^{(2)}$ are the first-order two-body corrections where $P^M_{(m,n)}$ projects the $m^{th}$ and $n^{th}$ particle onto a two-body state of relative angular momentum $M$, and $V_M^{(3)}$ are the lowest order three-body corrections with $P^M_{(m,n,o)}$ the projection of the $m^{th},$ $n^{th},$ and $o^{th}$ particles onto a three-body state of relative angular momentum $M$ ~\onlinecite{AngularStates}.  All of the potential terms depend upon the well width $w$.  We measure the energy in units of $e^2/\epsilon\ell_B$ and lengths in units of $\ell_B=\sqrt{\hbar c/eB}$.

We diagonalize Eq. (\ref{hamiltonian}) by breaking up the Hilbert space into subspaces as the rotational invariance causes the subspaces of fixed total angular momentum to decouple, so they are diagonalized individually \cite{5/2Disk}.  From this, we extract the lowest energy state of each subspace and consider the state with the lowest global energy to be the ground state at that $d$ and $\kappa.$  We use the total angular momentum of the Pfaffian, $N(2N-3)/2,$ $N$ is the number of electrons, and of the anti-Pfaffian to be $S(S-1)/2-(S-N)[2(S-N)-3]/2,$ $S$ is the number of available states, to identify regions where these states may be realized.

\begin{figure}
% GNUPLOT: LaTeX picture with Postscript
\begingroup
  \makeatletter
  \providecommand\color[2][]{%
    \GenericError{(gnuplot) \space\space\space\@spaces}{%
      Package color not loaded in conjunction with
      terminal option `colourtext'%
    }{See the gnuplot documentation for explanation.%
    }{Either use 'blacktext' in gnuplot or load the package
      color.sty in LaTeX.}%
    \renewcommand\color[2][]{}%
  }%
  \providecommand\includegraphics[2][]{%
    \GenericError{(gnuplot) \space\space\space\@spaces}{%
      Package graphicx or graphics not loaded%
    }{See the gnuplot documentation for explanation.%
    }{The gnuplot epslatex terminal needs graphicx.sty or graphics.sty.}%
    \renewcommand\includegraphics[2][]{}%
  }%
  \providecommand\rotatebox[2]{#2}%
  \@ifundefined{ifGPcolor}{%
    \newif\ifGPcolor
    \GPcolorfalse
  }{}%
  \@ifundefined{ifGPblacktext}{%
    \newif\ifGPblacktext
    \GPblacktexttrue
  }{}%
  % define a \g@addto@macro without @ in the name:
  \let\gplgaddtomacro\g@addto@macro
  % define empty templates for all commands taking text:
  \gdef\gplbacktext{}%
  \gdef\gplfronttext{}%
  \makeatother
  \ifGPblacktext
    % no textcolor at all
    \def\colorrgb#1{}%
    \def\colorgray#1{}%
  \else
    % gray or color?
    \ifGPcolor
      \def\colorrgb#1{\color[rgb]{#1}}%
      \def\colorgray#1{\color[gray]{#1}}%
      \expandafter\def\csname LTw\endcsname{\color{white}}%
      \expandafter\def\csname LTb\endcsname{\color{black}}%
      \expandafter\def\csname LTa\endcsname{\color{black}}%
      \expandafter\def\csname LT0\endcsname{\color[rgb]{1,0,0}}%
      \expandafter\def\csname LT1\endcsname{\color[rgb]{0,1,0}}%
      \expandafter\def\csname LT2\endcsname{\color[rgb]{0,0,1}}%
      \expandafter\def\csname LT3\endcsname{\color[rgb]{1,0,1}}%
      \expandafter\def\csname LT4\endcsname{\color[rgb]{0,1,1}}%
      \expandafter\def\csname LT5\endcsname{\color[rgb]{1,1,0}}%
      \expandafter\def\csname LT6\endcsname{\color[rgb]{0,0,0}}%
      \expandafter\def\csname LT7\endcsname{\color[rgb]{1,0.3,0}}%
      \expandafter\def\csname LT8\endcsname{\color[rgb]{0.5,0.5,0.5}}%
    \else
      % gray
      \def\colorrgb#1{\color{black}}%
      \def\colorgray#1{\color[gray]{#1}}%
      \expandafter\def\csname LTw\endcsname{\color{white}}%
      \expandafter\def\csname LTb\endcsname{\color{black}}%
      \expandafter\def\csname LTa\endcsname{\color{black}}%
      \expandafter\def\csname LT0\endcsname{\color{black}}%
      \expandafter\def\csname LT1\endcsname{\color{black}}%
      \expandafter\def\csname LT2\endcsname{\color{black}}%
      \expandafter\def\csname LT3\endcsname{\color{black}}%
      \expandafter\def\csname LT4\endcsname{\color{black}}%
      \expandafter\def\csname LT5\endcsname{\color{black}}%
      \expandafter\def\csname LT6\endcsname{\color{black}}%
      \expandafter\def\csname LT7\endcsname{\color{black}}%
      \expandafter\def\csname LT8\endcsname{\color{black}}%
    \fi
  \fi
  \setlength{\unitlength}{0.0500bp}%
  \begin{picture}(4648.00,3854.00)%
    \gplgaddtomacro\gplbacktext{%
      \csname LTb\endcsname%
      \put(734,860){\makebox(0,0)[r]{\strut{} 0}}%
      \put(734,1041){\makebox(0,0)[r]{\strut{} 0.1}}%
      \put(734,1222){\makebox(0,0)[r]{\strut{} 0.2}}%
      \put(734,1402){\makebox(0,0)[r]{\strut{} 0.3}}%
      \put(734,1583){\makebox(0,0)[r]{\strut{} 0.4}}%
      \put(734,1763){\makebox(0,0)[r]{\strut{} 0.5}}%
      \put(734,1944){\makebox(0,0)[r]{\strut{} 0.6}}%
      \put(734,2125){\makebox(0,0)[r]{\strut{} 0.7}}%
      \put(734,2305){\makebox(0,0)[r]{\strut{} 0.8}}%
      \put(734,2486){\makebox(0,0)[r]{\strut{} 0.9}}%
      \put(734,2667){\makebox(0,0)[r]{\strut{} 1}}%
      \put(734,2847){\makebox(0,0)[r]{\strut{} 1.1}}%
      \put(734,3028){\makebox(0,0)[r]{\strut{} 1.2}}%
      \put(734,3208){\makebox(0,0)[r]{\strut{} 1.3}}%
      \put(734,3389){\makebox(0,0)[r]{\strut{} 1.4}}%
      \put(734,3570){\makebox(0,0)[r]{\strut{} 1.5}}%
      \put(1008,487){\makebox(0,0){\strut{} 0}}%
      \put(1325,487){\makebox(0,0){\strut{} 0.2}}%
      \put(1642,487){\makebox(0,0){\strut{} 0.4}}%
      \put(1959,487){\makebox(0,0){\strut{} 0.6}}%
      \put(2276,487){\makebox(0,0){\strut{} 0.8}}%
      \put(2593,487){\makebox(0,0){\strut{} 1}}%
      \put(100,2215){\rotatebox{-270}{\makebox(0,0){\strut{}$d \left( \ell_B \right)$}}}%
      \put(1800,157){\makebox(0,0){\strut{}$\kappa$}}%
      \colorrgb{0.00,0.00,0.00}%
      \put(1008,1583){\makebox(0,0)[l]{\strut{}Unuseable}}%
    }%
    \gplgaddtomacro\gplfronttext{%
      \colorrgb{1.00,1.00,1.00}%
      \put(2333,1059){\makebox(0,0)[l]{\strut{}\textcolor{white}{(a)}}}%
      \colorrgb{0.00,0.00,0.00}%
      \put(1008,2958){\rotatebox{-270}{\makebox(0,0)[l]{\strut{}$M=101$}}}%
      \put(1325,2386){\rotatebox{-270}{\makebox(0,0)[l]{\strut{}$M=85$}}}%
      \put(1325,1402){\rotatebox{-270}{\makebox(0,0)[l]{\strut{}$M=75$}}}%
      \put(1642,1402){\rotatebox{-270}{\makebox(0,0)[l]{\strut{}$M=65$}}}%
      \put(1800,1222){\rotatebox{-270}{\makebox(0,0)[l]{\strut{}$M=55$}}}%
      \colorrgb{1.00,1.00,1.00}%
      \put(2434,2125){\rotatebox{-270}{\textcolor{white}{\makebox(0,0)[l]{\strut{}$M=45$}}}}
    }%
    \gplgaddtomacro\gplbacktext{%
      \csname LTb\endcsname%
      \put(2751,487){\makebox(0,0){\strut{} 0}}%
      \put(3068,487){\makebox(0,0){\strut{} 0.2}}%
      \put(3385,487){\makebox(0,0){\strut{} 0.4}}%
      \put(3701,487){\makebox(0,0){\strut{} 0.6}}%
      \put(4018,487){\makebox(0,0){\strut{} 0.8}}%
      \put(4335,487){\makebox(0,0){\strut{} 1}}%
      \put(3543,157){\makebox(0,0){\strut{}$\kappa$}}%
    }%
    \gplgaddtomacro\gplfronttext{%
      \colorrgb{1.00,1.00,1.00}%
      \put(4075,1059){\makebox(0,0)[l]{\strut{}\textcolor{white}{(b)}}}%
      \colorrgb{0.00,0.00,0.00}%
      \put(2910,2667){\rotatebox{-270}{\makebox(0,0)[l]{\strut{}$M=101$}}}%
      \put(3226,2225){\rotatebox{-270}{\makebox(0,0)[l]{\strut{}$M=85$}}}%
      \put(3385,1763){\rotatebox{-270}{\makebox(0,0)[l]{\strut{}$M=75$}}}%
      \put(3543,1583){\rotatebox{-270}{\makebox(0,0)[l]{\strut{}$M=65$}}}%
      \put(3701,1222){\rotatebox{-270}{\makebox(0,0)[l]{\strut{}$M=55$}}}%
      \colorrgb{1.00,1.00,1.00}%
      \put(4176,2125){\rotatebox{-270}{\textcolor{white}{\makebox(0,0)[l]{\strut{}$M=45$}}}}%
    }%
    \gplbacktext
    \put(0,0){\includegraphics{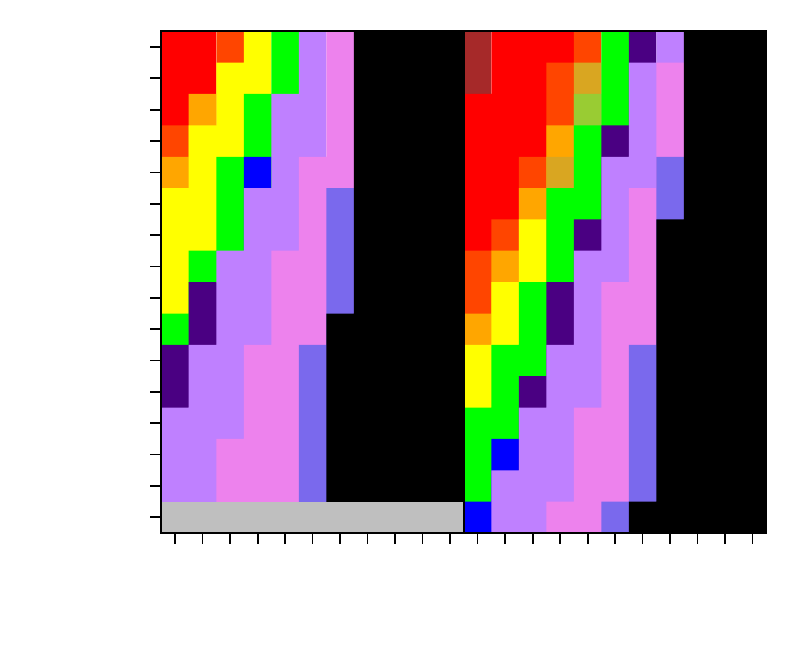}}%
    \gplfronttext
  \end{picture}%
\endgroup
\caption{\label{Phase} (a) The phase diagram of 10 particles in 18 states with $w=0\ell_B$ as the distance to the neutralizing disk $d$ and the LL mixing strength $\kappa$ are varied. (b) The phase diagram for varying $d$ and $\kappa$ with $w=1\ell_B$.   The potential Pfaffian region is highlighted in green for both diagrams while the anti-Pfaffian is highlighted in red.  It is easy to see that the inclusion of the finite well thickness expands both of these regions.}
\end{figure}

The results of this procedure are shown in Fig.1.  The noticeable effect of the LL mixing is that all of the observed states arise at larger $d$ than in the absence of mixing.  Thus, the LL mixing strength and the confinement by the neutralizing disk, due to the generally attractive nature of the LL mixing terms, balances with the Coulomb repulsion.  As $d$ increases, the confinement of the EG by the neutralizing background weakens and the Coulomb interaction pushes the EG towards the edge, while the LL mixing interaction can pull the electrons back to the center as it is increased. Thus, from the model simulation, larger $d$ correspond to larger $\kappa$.

Due to the small size of our system, with a neutralizing disk of radius $\sim 6\ell_B$, we do not exceed $d=1.5\ell_B$ to maintain a charge distribution similar to experiment, as in previous simulations on the disk \cite{5/2Disk}.  Realistic experimental separations are significantly larger, being closer to $\sim 10\ell_B$.  As larger $d$ leads to weaker confinement by the neutralizing background, $\kappa$ must be increased to compensate and realize the same state as was observed at lower $d$.  Thus, from the obtained relation between $d$ and $\kappa$, we expect the larger value of $\kappa$ in the appropriate interval of values, in larger systems.  Therefore, both $d$ and $\kappa$ are vital for the realization of the non-abelian states in experimentally relevant ranges of disk separations and LL mixing strength.

The anti-Pfaffian $M=101$ region favors a smoother edge (large $d$) and strong magnetic fields (small $\kappa$), while the Pfaffian $M=85$ region favors small $d$ and large $\kappa.$  This is the opposite of observations in Ref. \onlinecite{ED4} where the Pfaffian appears at $\kappa<\kappa_c\left(w\right)$ and transitions to the anti-Pfaffian when $\kappa>\kappa_c\left(w\right).$  This is a result of the interaction with the neutralizing background as the overlap with the anti-Pfaffian increases with increasing $\kappa$ and the overlap with the Pfaffian falls with increasing $\kappa$.  We suggest that transitions observed in experiment \cite{Transition} follow the form in Fig. \ref{Phase}.

Another noticeable feature is the $M=45$ collapsed state.  This region represents the collapse of the electrons to the center of the disk where they are supported only by degeneracy pressure.  This is a result of LL mixing and the neutralizing disk potential overcoming Coulomb repulsion entirely.  Comparing our 10-particle calculations to calculations with 8 particles, we see that this state is pushed to higher $\kappa$ by the introduction of new compressible stripe states as the particle number increases.

These key points distinguish our results for the 2D phase diagram from what is expected from Ref. \onlinecite{5/2Disk}. When the finite thickness of the well is introduced, the potentially incompressible states appear much stronger than in the 2D case, with the incompressible states occurring at much lower $d$ than in the 2D case.  However, several features of both cases do not differ that dramatically, particularly the presence of the $M=101$ region and the $M=85$ region separated by a series of compressible stripe states, with the $M=85$ region having a strong overlap with the Pfaffian throughout.

\begin{figure}
% GNUPLOT: LaTeX picture with Postscript
\begingroup
  \makeatletter
  \providecommand\color[2][]{%
    \GenericError{(gnuplot) \space\space\space\@spaces}{%
      Package color not loaded in conjunction with
      terminal option `colourtext'%
    }{See the gnuplot documentation for explanation.%
    }{Either use 'blacktext' in gnuplot or load the package
      color.sty in LaTeX.}%
    \renewcommand\color[2][]{}%
  }%
  \providecommand\includegraphics[2][]{%
    \GenericError{(gnuplot) \space\space\space\@spaces}{%
      Package graphicx or graphics not loaded%
    }{See the gnuplot documentation for explanation.%
    }{The gnuplot epslatex terminal needs graphicx.sty or graphics.sty.}%
    \renewcommand\includegraphics[2][]{}%
  }%
  \providecommand\rotatebox[2]{#2}%
  \@ifundefined{ifGPcolor}{%
    \newif\ifGPcolor
    \GPcolorfalse
  }{}%
  \@ifundefined{ifGPblacktext}{%
    \newif\ifGPblacktext
    \GPblacktexttrue
  }{}%
  % define a \g@addto@macro without @ in the name:
  \let\gplgaddtomacro\g@addto@macro
  % define empty templates for all commands taking text:
  \gdef\gplbacktext{}%
  \gdef\gplfronttext{}%
  \makeatother
  \ifGPblacktext
    % no textcolor at all
    \def\colorrgb#1{}%
    \def\colorgray#1{}%
  \else
    % gray or color?
    \ifGPcolor
      \def\colorrgb#1{\color[rgb]{#1}}%
      \def\colorgray#1{\color[gray]{#1}}%
      \expandafter\def\csname LTw\endcsname{\color{white}}%
      \expandafter\def\csname LTb\endcsname{\color{black}}%
      \expandafter\def\csname LTa\endcsname{\color{black}}%
      \expandafter\def\csname LT0\endcsname{\color[rgb]{1,0,0}}%
      \expandafter\def\csname LT1\endcsname{\color[rgb]{0,1,0}}%
      \expandafter\def\csname LT2\endcsname{\color[rgb]{0,0,1}}%
      \expandafter\def\csname LT3\endcsname{\color[rgb]{1,0,1}}%
      \expandafter\def\csname LT4\endcsname{\color[rgb]{0,1,1}}%
      \expandafter\def\csname LT5\endcsname{\color[rgb]{1,1,0}}%
      \expandafter\def\csname LT6\endcsname{\color[rgb]{0,0,0}}%
      \expandafter\def\csname LT7\endcsname{\color[rgb]{1,0.3,0}}%
      \expandafter\def\csname LT8\endcsname{\color[rgb]{0.5,0.5,0.5}}%
    \else
      % gray
      \def\colorrgb#1{\color{black}}%
      \def\colorgray#1{\color[gray]{#1}}%
      \expandafter\def\csname LTw\endcsname{\color{white}}%
      \expandafter\def\csname LTb\endcsname{\color{black}}%
      \expandafter\def\csname LTa\endcsname{\color{black}}%
      \expandafter\def\csname LT0\endcsname{\color{black}}%
      \expandafter\def\csname LT1\endcsname{\color{black}}%
      \expandafter\def\csname LT2\endcsname{\color{black}}%
      \expandafter\def\csname LT3\endcsname{\color{black}}%
      \expandafter\def\csname LT4\endcsname{\color{black}}%
      \expandafter\def\csname LT5\endcsname{\color{black}}%
      \expandafter\def\csname LT6\endcsname{\color{black}}%
      \expandafter\def\csname LT7\endcsname{\color{black}}%
      \expandafter\def\csname LT8\endcsname{\color{black}}%
    \fi
  \fi
  \setlength{\unitlength}{0.0500bp}%
  \begin{picture}(4648.00,4648.00)%
    \gplgaddtomacro\gplbacktext{%
      \csname LTb\endcsname%
      \put(734,3335){\makebox(0,0)[r]{\strut{} 0.01}}%
      \put(734,3551){\makebox(0,0)[r]{\strut{} 0.02}}%
      \put(734,3767){\makebox(0,0)[r]{\strut{} 0.03}}%
      \put(734,3983){\makebox(0,0)[r]{\strut{} 0.04}}%
      \put(734,4198){\makebox(0,0)[r]{\strut{} 0.05}}%
      \put(734,4414){\makebox(0,0)[r]{\strut{} 0.06}}%
      \put(100,2500){\rotatebox{-270}{\makebox(0,0){\strut{}$\Delta E \left( \frac{e^2}{ \epsilon \ell_B} \right)$}}}%
    }%
    \gplgaddtomacro\gplfronttext{%
      \csname LTb\endcsname%
      \put(2325,3230){\makebox(0,0)[l]{\strut{}(a)}}%
      \put(1124,3767){\rotatebox{-270}{\makebox(0,0)[l]{\strut{}\textbf{90}}}}%
      \put(1498,3767){\rotatebox{-270}{\makebox(0,0)[l]{\strut{}\textbf{85}}}}%
      \put(1986,3767){\rotatebox{-270}{\makebox(0,0)[l]{\strut{}\textbf{75}}}}%
      \put(2425,3767){\rotatebox{-270}{\makebox(0,0)[l]{\strut{}\textbf{65}}}}%
    }%
    \gplgaddtomacro\gplbacktext{%
    }%
    \gplgaddtomacro\gplfronttext{%
      \csname LTb\endcsname%
      \put(4184,3230){\makebox(0,0)[l]{\strut{}(b)}}%
      \put(2951,3767){\rotatebox{-270}{\makebox(0,0)[l]{\strut{}\textbf{101}}}}%
      \put(3243,3767){\rotatebox{-270}{\makebox(0,0)[l]{\strut{}\textbf{95}}}}%
      \put(3540,3767){\rotatebox{-270}{\makebox(0,0)[l]{\strut{}\textbf{91}}}}%
      \put(3829,3767){\rotatebox{-270}{\makebox(0,0)[l]{\strut{}\textbf{85}}}}%
      \put(4089,3767){\rotatebox{-270}{\makebox(0,0)[l]{\strut{}\textbf{80}}}}%
      \put(4251,3767){\rotatebox{-270}{\makebox(0,0)[l]{\strut{}\textbf{75}}}}%
    }%
    \gplgaddtomacro\gplbacktext{%
      \csname LTb\endcsname%
      \put(734,2018){\makebox(0,0)[r]{\strut{} 0.01}}%
      \put(734,2234){\makebox(0,0)[r]{\strut{} 0.02}}%
      \put(734,2450){\makebox(0,0)[r]{\strut{} 0.03}}%
      \put(734,2666){\makebox(0,0)[r]{\strut{} 0.04}}%
      \put(734,2881){\makebox(0,0)[r]{\strut{} 0.05}}%
      \put(734,3097){\makebox(0,0)[r]{\strut{} 0.06}}%
      \put(-168,2439){\rotatebox{-270}{\makebox(0,0){\strut{} }}}%
    }%
    \gplgaddtomacro\gplfronttext{%
      \csname LTb\endcsname%
      \put(2325,1913){\makebox(0,0)[l]{\strut{}(c)}}%
      \put(1154,2417){\rotatebox{-270}{\makebox(0,0)[l]{\strut{}\textbf{90}}}}%
      \put(1548,2417){\rotatebox{-270}{\makebox(0,0)[l]{\strut{}\textbf{85}}}}%
      \put(2036,2417){\rotatebox{-270}{\makebox(0,0)[l]{\strut{}\textbf{75}}}}%
      \put(2475,2417){\rotatebox{-270}{\makebox(0,0)[l]{\strut{}\textbf{65}}}}%
    }%
    \gplgaddtomacro\gplbacktext{%
    }%
    \gplgaddtomacro\gplfronttext{%
      \csname LTb\endcsname%
      \put(4184,1913){\makebox(0,0)[l]{\strut{}(d)}}%
      \put(3243,2417){\rotatebox{-270}{\makebox(0,0)[l]{\strut{}\textbf{99}}}}%
      \put(3740,2417){\rotatebox{-270}{\makebox(0,0)[l]{\strut{}\textbf{92}}}}%
      \put(4079,2417){\rotatebox{-270}{\makebox(0,0)[l]{\strut{}\textbf{85}}}}%
    }%
    \gplgaddtomacro\gplbacktext{%
      \csname LTb\endcsname%
      \put(734,701){\makebox(0,0)[r]{\strut{} 0.01}}%
      \put(734,917){\makebox(0,0)[r]{\strut{} 0.02}}%
      \put(734,1133){\makebox(0,0)[r]{\strut{} 0.03}}%
      \put(734,1349){\makebox(0,0)[r]{\strut{} 0.04}}%
      \put(734,1565){\makebox(0,0)[r]{\strut{} 0.05}}%
      \put(734,1781){\makebox(0,0)[r]{\strut{} 0.06}}%
      \put(929,181){\makebox(0,0){\strut{} 0.1}}%
      \put(1254,181){\makebox(0,0){\strut{} 0.2}}%
      \put(1579,181){\makebox(0,0){\strut{} 0.3}}%
      \put(1905,181){\makebox(0,0){\strut{} 0.4}}%
      \put(2230,181){\makebox(0,0){\strut{} 0.5}}%
      \put(2555,181){\makebox(0,0){\strut{} 0.6}}%
      \put(-168,1122){\rotatebox{-270}{\makebox(0,0){\strut{} }}}%
      \put(1742,0){\makebox(0,0){\strut{}$\kappa$}}%
    }%
    \gplgaddtomacro\gplfronttext{%
      \csname LTb\endcsname%
      \put(2325,596){\makebox(0,0)[l]{\strut{}(e)}}%
      \put(1204,1157){\rotatebox{-270}{\makebox(0,0)[l]{\strut{}\textbf{90}}}}%
      \put(1648,1157){\rotatebox{-270}{\makebox(0,0)[l]{\strut{}\textbf{85}}}}%
      \put(2136,1157){\rotatebox{-270}{\makebox(0,0)[l]{\strut{}\textbf{75}}}}%
      \put(2495,1157){\rotatebox{-270}{\makebox(0,0)[l]{\strut{}\textbf{65}}}}%
    }%
    \gplgaddtomacro\gplbacktext{%
      \csname LTb\endcsname%
      \put(2788,181){\makebox(0,0){\strut{} 0.1}}%
      \put(3113,181){\makebox(0,0){\strut{} 0.2}}%
      \put(3438,181){\makebox(0,0){\strut{} 0.3}}%
      \put(3764,181){\makebox(0,0){\strut{} 0.4}}%
      \put(4089,181){\makebox(0,0){\strut{} 0.5}}%
      \put(4414,181){\makebox(0,0){\strut{} 0.6}}%
      \put(3601,0){\makebox(0,0){\strut{}$\kappa$}}%
    }%
    \gplgaddtomacro\gplfronttext{%
      \csname LTb\endcsname%
      \put(4184,596){\makebox(0,0)[l]{\strut{}(f)}}%
      \put(2933,1157){\rotatebox{-270}{\makebox(0,0)[l]{\strut{}\textbf{108}}}}%
      \put(3393,1157){\rotatebox{-270}{\makebox(0,0)[l]{\strut{}\textbf{99}}}}%
      \put(3840,1157){\rotatebox{-270}{\makebox(0,0)[l]{\strut{}\textbf{95}}}}%
      \put(4279,1157){\rotatebox{-270}{\makebox(0,0)[l]{\strut{}\textbf{85}}}}%
    }%
    \gplbacktext
    \put(0,0){\includegraphics{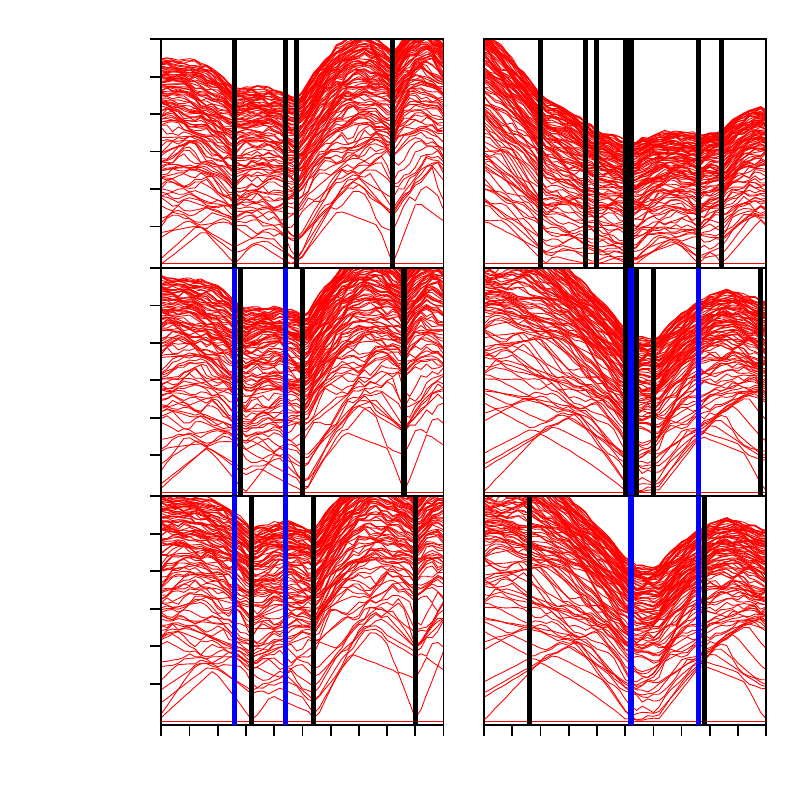}}%
    \gplfronttext
  \end{picture}%
\endgroup
\caption{\label{12ZQuasi} The 100 lowest energy levels of 10 particles in 18 states for $d=1.2\ell_B$ with $w=0\ell_B$ (left) and $w=1\ell_B$l (right). The individual graphs relate to different $U$ for the Gaussian tip potential of Eq. (\ref{gaussian}) with $\sigma\simeq4$: (a),(b) No Gaussian tip potential, (c) $U=0.01e^2/\epsilon\ell_B$, (d) $U=0.05e^2/\epsilon\ell_B$, (e) $U=0.02e^2/\epsilon\ell_B$, (f) $U=0.1e^2/\epsilon\ell_B$. The locations of the phase transitions are highlighted by the vertical black bars and the $\kappa$ range of the $U=0$ $M=85$ phase is bordered by blue in all plots.}
\end{figure}

Analyzing the phase transitions, we take a cut across the phase diagram at fixed $d=1.2\ell_B$, performing a  higher resolution sweep of the LL mixing strength as shown in Fig. \ref{12ZQuasi}a,b.   A striking difference between the pure 2D case $w=0$ and the finite width confinement, $w=1\ell_B$, immediately becomes apparent as we look at the energy gaps.  In the system with $w=0\ell_B,$  all ground states develop an energy gap, as we move away from the phase transition.  For $w=1\ell_B,$ the energy gaps for the stripe phases have closed and only the candidate incompressible states continue to possess an energy gap.  In the $M=101$ region, we take this and the rapid increase of the overlap integral as $\kappa$ and $d$ increase as an indication of this region belonging to the same class as the anti-Pfaffian.

With this detailed cut, we also explore the charge $e/4$ quasihole properties of this phase by modeling  the effects of a potential \cite{5/2Disk}
\begin{equation}
\label{gaussian}
\hat{H}_U=U\sum_m\exp\frac{-m^2}{2\sigma^2}a^\dagger_ma_m,
\end{equation}
equivalent to applying a repulsive Gaussian tip to the center of the disk. $U$ is the strength and $\sigma$ is the width of the potential, which correlates with the size of the quasihole. The results are shown for Fig. \ref{12ZQuasi}c-f, alternating between $w=0\ell_B$ and $w=1\ell_B$.

As we increase the strength of the tip potential, for $w=0\ell_B$, the $M=85$ phase begins to be displaced by the neighboring $M=90$ phase associated with the formation of a charge $e/4$ quasihole \cite{Quasiholes}.  In previous work on the disk, such quasihole states were introduced using a tip potential with a size $\sigma\simeq3$, but in order to introduce a quasihole excitation here, we must increase the size of the Gaussian tip to $\sigma\simeq4$.  For $w=1\ell_B$, we are no longer able to introduce a single quasihole excitation, but we may introduce a pair of quasiholes at much stronger strengths.  This may be an indication of pairing in the ground state as the two-quasihole state does not change the boundary conditions and leaves the edge structure unchanged.

\begin{figure}
% GNUPLOT: LaTeX picture with Postscript
\begingroup
  \makeatletter
  \providecommand\color[2][]{%
    \GenericError{(gnuplot) \space\space\space\@spaces}{%
      Package color not loaded in conjunction with
      terminal option `colourtext'%
    }{See the gnuplot documentation for explanation.%
    }{Either use 'blacktext' in gnuplot or load the package
      color.sty in LaTeX.}%
    \renewcommand\color[2][]{}%
  }%
  \providecommand\includegraphics[2][]{%
    \GenericError{(gnuplot) \space\space\space\@spaces}{%
      Package graphicx or graphics not loaded%
    }{See the gnuplot documentation for explanation.%
    }{The gnuplot epslatex terminal needs graphicx.sty or graphics.sty.}%
    \renewcommand\includegraphics[2][]{}%
  }%
  \providecommand\rotatebox[2]{#2}%
  \@ifundefined{ifGPcolor}{%
    \newif\ifGPcolor
    \GPcolorfalse
  }{}%
  \@ifundefined{ifGPblacktext}{%
    \newif\ifGPblacktext
    \GPblacktexttrue
  }{}%
  % define a \g@addto@macro without @ in the name:
  \let\gplgaddtomacro\g@addto@macro
  % define empty templates for all commands taking text:
  \gdef\gplbacktext{}%
  \gdef\gplfronttext{}%
  \makeatother
  \ifGPblacktext
    % no textcolor at all
    \def\colorrgb#1{}%
    \def\colorgray#1{}%
  \else
    % gray or color?
    \ifGPcolor
      \def\colorrgb#1{\color[rgb]{#1}}%
      \def\colorgray#1{\color[gray]{#1}}%
      \expandafter\def\csname LTw\endcsname{\color{white}}%
      \expandafter\def\csname LTb\endcsname{\color{black}}%
      \expandafter\def\csname LTa\endcsname{\color{black}}%
      \expandafter\def\csname LT0\endcsname{\color[rgb]{1,0,0}}%
      \expandafter\def\csname LT1\endcsname{\color[rgb]{0,1,0}}%
      \expandafter\def\csname LT2\endcsname{\color[rgb]{0,0,1}}%
      \expandafter\def\csname LT3\endcsname{\color[rgb]{1,0,1}}%
      \expandafter\def\csname LT4\endcsname{\color[rgb]{0,1,1}}%
      \expandafter\def\csname LT5\endcsname{\color[rgb]{1,1,0}}%
      \expandafter\def\csname LT6\endcsname{\color[rgb]{0,0,0}}%
      \expandafter\def\csname LT7\endcsname{\color[rgb]{1,0.3,0}}%
      \expandafter\def\csname LT8\endcsname{\color[rgb]{0.5,0.5,0.5}}%
    \else
      % gray
      \def\colorrgb#1{\color{black}}%
      \def\colorgray#1{\color[gray]{#1}}%
      \expandafter\def\csname LTw\endcsname{\color{white}}%
      \expandafter\def\csname LTb\endcsname{\color{black}}%
      \expandafter\def\csname LTa\endcsname{\color{black}}%
      \expandafter\def\csname LT0\endcsname{\color{black}}%
      \expandafter\def\csname LT1\endcsname{\color{black}}%
      \expandafter\def\csname LT2\endcsname{\color{black}}%
      \expandafter\def\csname LT3\endcsname{\color{black}}%
      \expandafter\def\csname LT4\endcsname{\color{black}}%
      \expandafter\def\csname LT5\endcsname{\color{black}}%
      \expandafter\def\csname LT6\endcsname{\color{black}}%
      \expandafter\def\csname LT7\endcsname{\color{black}}%
      \expandafter\def\csname LT8\endcsname{\color{black}}%
    \fi
  \fi
  \setlength{\unitlength}{0.0500bp}%
  \begin{picture}(4648.00,4648.00)%
    \gplgaddtomacro\gplbacktext{%
      \csname LTb\endcsname%
      \put(1078,3335){\makebox(0,0)[r]{\strut{} 0.01}}%
      \put(1078,3551){\makebox(0,0)[r]{\strut{} 0.02}}%
      \put(1078,3767){\makebox(0,0)[r]{\strut{} 0.03}}%
      \put(1078,3983){\makebox(0,0)[r]{\strut{} 0.04}}%
      \put(1078,4198){\makebox(0,0)[r]{\strut{} 0.05}}%
      \put(1078,4414){\makebox(0,0)[r]{\strut{} 0.06}}%
      \put(100,2500){\rotatebox{-270}{\makebox(0,0){\strut{}$\Delta E \left( \frac{e^2}{ \epsilon \ell_B} \right)$}}}%
    }%
    \gplgaddtomacro\gplfronttext{%
      \csname LTb\endcsname%
      \put(1571,3659){\rotatebox{-270}{\makebox(0,0)[l]{\strut{}\textbf{101}}}}%
      \put(2107,3659){\rotatebox{-270}{\makebox(0,0)[l]{\strut{}\textbf{95}}}}%
      \put(2405,3659){\rotatebox{-270}{\makebox(0,0)[l]{\strut{}\textbf{93}}}}%
      \put(2613,3659){\rotatebox{-270}{\makebox(0,0)[l]{\strut{}\textbf{91}}}}%
      \put(3179,3659){\rotatebox{-270}{\makebox(0,0)[l]{\strut{}\textbf{85}}}}%
      \put(3655,3659){\rotatebox{-270}{\makebox(0,0)[l]{\strut{}\textbf{80}}}}%
      \put(4053,3659){\rotatebox{-270}{\makebox(0,0)[l]{\strut{}\textbf{75}}}}%
      \put(4013,3233){\makebox(0,0)[l]{\strut{}(a)}}%
    }%
    \gplgaddtomacro\gplbacktext{%
      \csname LTb\endcsname%
      \put(1078,2018){\makebox(0,0)[r]{\strut{} 0.01}}%
      \put(1078,2234){\makebox(0,0)[r]{\strut{} 0.02}}%
      \put(1078,2450){\makebox(0,0)[r]{\strut{} 0.03}}%
      \put(1078,2666){\makebox(0,0)[r]{\strut{} 0.04}}%
      \put(1078,2881){\makebox(0,0)[r]{\strut{} 0.05}}%
      \put(1078,3097){\makebox(0,0)[r]{\strut{} 0.06}}%
      \put(176,2439){\rotatebox{-270}{\makebox(0,0){\strut{} }}}%
    }%
    \gplgaddtomacro\gplfronttext{%
      \csname LTb\endcsname%
      \put(1621,2339){\rotatebox{-270}{\makebox(0,0)[l]{\strut{}\textbf{101}}}}%
      \put(2307,2339){\rotatebox{-270}{\makebox(0,0)[l]{\strut{}\textbf{95}}}}%
      \put(2688,2339){\rotatebox{-270}{\makebox(0,0)[l]{\strut{}\textbf{91}}}}%
      \put(2838,2339){\rotatebox{-270}{\makebox(0,0)[l]{\strut{}\textbf{90}}}}%
      \put(3279,2339){\rotatebox{-270}{\makebox(0,0)[l]{\strut{}\textbf{85}}}}%
      \put(3725,2339){\rotatebox{-270}{\makebox(0,0)[l]{\strut{}\textbf{80}}}}%
      \put(4073,2339){\rotatebox{-270}{\makebox(0,0)[l]{\strut{}\textbf{75}}}}%
      \put(4013,1913){\makebox(0,0)[l]{\strut{}(b)}}%
    }%
    \gplgaddtomacro\gplbacktext{%
      \csname LTb\endcsname%
      \put(1078,701){\makebox(0,0)[r]{\strut{} 0.01}}%
      \put(1078,917){\makebox(0,0)[r]{\strut{} 0.02}}%
      \put(1078,1133){\makebox(0,0)[r]{\strut{} 0.03}}%
      \put(1078,1349){\makebox(0,0)[r]{\strut{} 0.04}}%
      \put(1078,1565){\makebox(0,0)[r]{\strut{} 0.05}}%
      \put(1078,1781){\makebox(0,0)[r]{\strut{} 0.06}}%
      \put(1273,181){\makebox(0,0){\strut{} 0.1}}%
      \put(1869,181){\makebox(0,0){\strut{} 0.2}}%
      \put(2464,181){\makebox(0,0){\strut{} 0.3}}%
      \put(3060,181){\makebox(0,0){\strut{} 0.4}}%
      \put(3655,181){\makebox(0,0){\strut{} 0.5}}%
      \put(4251,181){\makebox(0,0){\strut{} 0.6}}%
      \put(176,1122){\rotatebox{-270}{\makebox(0,0){\strut{} }}}%
      \put(2762,-148){\makebox(0,0){\strut{}$\kappa$}}%
    }%
    \gplgaddtomacro\gplfronttext{%
      \csname LTb\endcsname%
      \put(1471,1019){\rotatebox{-270}{\makebox(0,0)[l]{\strut{}\textbf{101}}}}%
      \put(1907,1019){\rotatebox{-270}{\makebox(0,0)[l]{\strut{}\textbf{99}}}}%
      \put(2448,1019){\rotatebox{-270}{\makebox(0,0)[l]{\strut{}\textbf{95}}}}%
      \put(2858,1019){\rotatebox{-270}{\makebox(0,0)[l]{\strut{}\textbf{90}}}}%
      \put(3479,1019){\rotatebox{-270}{\makebox(0,0)[l]{\strut{}\textbf{85}}}}%
      \put(4103,1019){\rotatebox{-270}{\makebox(0,0)[l]{\strut{}\textbf{75}}}}%
      \put(4013,596){\makebox(0,0)[l]{\strut{}(c)}}%
    }%
    \gplbacktext
    \put(0,0){\includegraphics{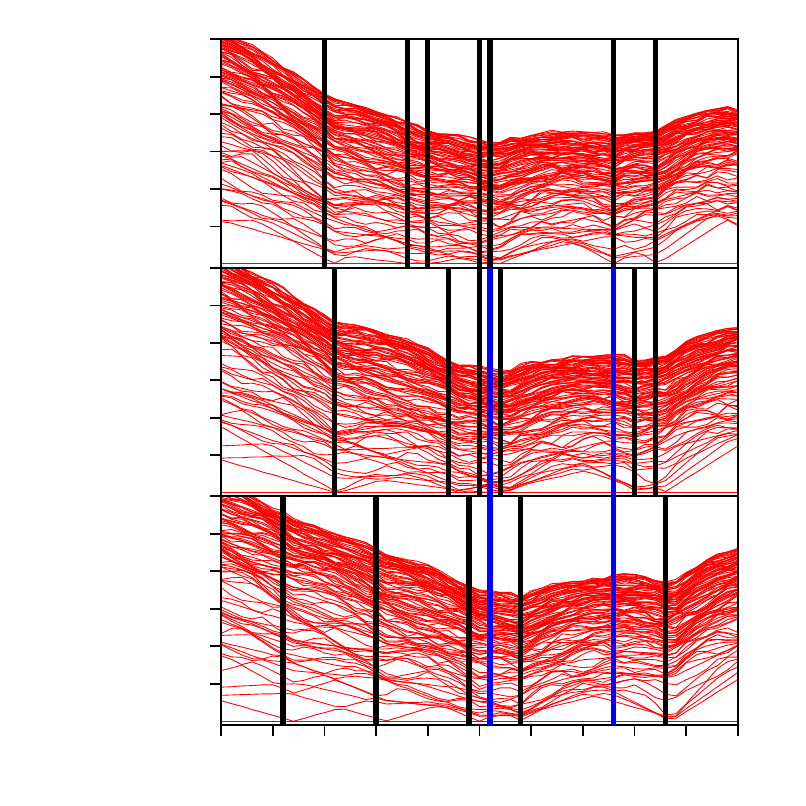}}%
    \gplfronttext
  \end{picture}%
\endgroup
\caption{\label{12ZQuasiBigger}The same as in Fig. \ref{12ZQuasi}, but only results for $w=1\ell_B$ are shown with a Gaussian tip potential with $\sigma\simeq4.5$ and (a) $U=0,$ (b) $U=0.01e^2/\epsilon\ell_B$, (c) $U=0.02e^2/\epsilon\ell_B$.  The increased tip size creates and expands a phase with an $M=90$  ground state with a charge $e/4$ quasihole isolated at the center of the disk.}
\end{figure}

In order to understand the increase in size of the charge $e/4$ quasihole, we consider the classical effects of introducing LL mixing. Adiabatically turning on LL mixing by increasing $\kappa$, is formally equivalent to decreasing the magnetic field. Then a small amount of negative charge is transported from the center of the disk to the edge by this process and the equilibrium state of the same $M$ has charge located on the edge. As the charge $e/4$ quasiholes are a similar center of rotation, a similar effect should occur, increasing the size of the region of depleted charge. The introduction of finite thickness effects results in the weakening of the confining potential. The Coulomb repulsion requires stronger LL mixing to realize the $M=85$ state and so the charge $e/4$ quasiholes should be of a larger size. Thus, when $\sigma$ in Eq. (\ref{gaussian}) is increased, we expect that a single quasihole excitation in the $w=1\ell_B$ system will be introduced. The results for $\sigma\simeq4.5$ and $w=1\ell_B$ are shown in Fig. \ref{12ZQuasiBigger}. As $U$ is increased, a new $M=90$ state appears and begins to intrude into the region previously occupied by the $M=85$ Pfaffian state, just as in the $w=0\ell_B$ case.

\begin{figure}
% GNUPLOT: LaTeX picture with Postscript
\begingroup
  \makeatletter
  \providecommand\color[2][]{%
    \GenericError{(gnuplot) \space\space\space\@spaces}{%
      Package color not loaded in conjunction with
      terminal option `colourtext'%
    }{See the gnuplot documentation for explanation.%
    }{Either use 'blacktext' in gnuplot or load the package
      color.sty in LaTeX.}%
    \renewcommand\color[2][]{}%
  }%
  \providecommand\includegraphics[2][]{%
    \GenericError{(gnuplot) \space\space\space\@spaces}{%
      Package graphicx or graphics not loaded%
    }{See the gnuplot documentation for explanation.%
    }{The gnuplot epslatex terminal needs graphicx.sty or graphics.sty.}%
    \renewcommand\includegraphics[2][]{}%
  }%
  \providecommand\rotatebox[2]{#2}%
  \@ifundefined{ifGPcolor}{%
    \newif\ifGPcolor
    \GPcolorfalse
  }{}%
  \@ifundefined{ifGPblacktext}{%
    \newif\ifGPblacktext
    \GPblacktexttrue
  }{}%
  % define a \g@addto@macro without @ in the name:
  \let\gplgaddtomacro\g@addto@macro
  % define empty templates for all commands taking text:
  \gdef\gplbacktext{}%
  \gdef\gplfronttext{}%
  \makeatother
  \ifGPblacktext
    % no textcolor at all
    \def\colorrgb#1{}%
    \def\colorgray#1{}%
  \else
    % gray or color?
    \ifGPcolor
      \def\colorrgb#1{\color[rgb]{#1}}%
      \def\colorgray#1{\color[gray]{#1}}%
      \expandafter\def\csname LTw\endcsname{\color{white}}%
      \expandafter\def\csname LTb\endcsname{\color{black}}%
      \expandafter\def\csname LTa\endcsname{\color{black}}%
      \expandafter\def\csname LT0\endcsname{\color[rgb]{1,0,0}}%
      \expandafter\def\csname LT1\endcsname{\color[rgb]{0,1,0}}%
      \expandafter\def\csname LT2\endcsname{\color[rgb]{0,0,1}}%
      \expandafter\def\csname LT3\endcsname{\color[rgb]{1,0,1}}%
      \expandafter\def\csname LT4\endcsname{\color[rgb]{0,1,1}}%
      \expandafter\def\csname LT5\endcsname{\color[rgb]{1,1,0}}%
      \expandafter\def\csname LT6\endcsname{\color[rgb]{0,0,0}}%
      \expandafter\def\csname LT7\endcsname{\color[rgb]{1,0.3,0}}%
      \expandafter\def\csname LT8\endcsname{\color[rgb]{0.5,0.5,0.5}}%
    \else
      % gray
      \def\colorrgb#1{\color{black}}%
      \def\colorgray#1{\color[gray]{#1}}%
      \expandafter\def\csname LTw\endcsname{\color{white}}%
      \expandafter\def\csname LTb\endcsname{\color{black}}%
      \expandafter\def\csname LTa\endcsname{\color{black}}%
      \expandafter\def\csname LT0\endcsname{\color{black}}%
      \expandafter\def\csname LT1\endcsname{\color{black}}%
      \expandafter\def\csname LT2\endcsname{\color{black}}%
      \expandafter\def\csname LT3\endcsname{\color{black}}%
      \expandafter\def\csname LT4\endcsname{\color{black}}%
      \expandafter\def\csname LT5\endcsname{\color{black}}%
      \expandafter\def\csname LT6\endcsname{\color{black}}%
      \expandafter\def\csname LT7\endcsname{\color{black}}%
      \expandafter\def\csname LT8\endcsname{\color{black}}%
    \fi
  \fi
  \setlength{\unitlength}{0.0500bp}%
  \begin{picture}(4648.00,4648.00)%
    \gplgaddtomacro\gplbacktext{%
      \csname LTb\endcsname%
      \put(734,2455){\makebox(0,0)[r]{\strut{} 0}}%
      \put(734,2716){\makebox(0,0)[r]{\strut{} 0.02}}%
      \put(734,2977){\makebox(0,0)[r]{\strut{} 0.04}}%
      \put(734,3238){\makebox(0,0)[r]{\strut{} 0.06}}%
      \put(734,3500){\makebox(0,0)[r]{\strut{} 0.08}}%
      \put(734,3761){\makebox(0,0)[r]{\strut{} 0.1}}%
      \put(734,4022){\makebox(0,0)[r]{\strut{} 0.12}}%
      \put(734,4283){\makebox(0,0)[r]{\strut{} 0.14}}%
      \put(991,2041){\makebox(0,0){\strut{} 0}}%
      \put(1199,2041){\makebox(0,0){\strut{} 1}}%
      \put(1406,2041){\makebox(0,0){\strut{} 2}}%
      \put(1614,2041){\makebox(0,0){\strut{} 3}}%
      \put(1821,2041){\makebox(0,0){\strut{} 4}}%
      \put(2029,2041){\makebox(0,0){\strut{} 5}}%
      \put(100,3369){\rotatebox{-270}{\makebox(0,0){\strut{}$\Delta E \left( \frac{e^2}{\epsilon \ell_B} \right)$}}}%
      \put(1510,1711){\makebox(0,0){\strut{}$\Delta M$}}%
    }%
    \gplgaddtomacro\gplfronttext{%
      \csname LTb\endcsname%
      \put(1852,2429){\makebox(0,0)[l]{\strut{}(a)}}%
    }%
    \gplgaddtomacro\gplbacktext{%
      \put(2153,2041){\makebox(0,0){\strut{} 0}}%
      \put(2361,2041){\makebox(0,0){\strut{} 1}}%
      \put(2568,2041){\makebox(0,0){\strut{} 2}}%
      \put(2775,2041){\makebox(0,0){\strut{} 3}}%
      \put(2982,2041){\makebox(0,0){\strut{} 4}}%
      \put(3190,2041){\makebox(0,0){\strut{} 5}}%
      \put(2671,1711){\makebox(0,0){\strut{}$\Delta M$}}%
    }%
    \gplgaddtomacro\gplfronttext{%
      \csname LTb\endcsname%
      \put(3013,2429){\makebox(0,0)[l]{\strut{}(b)}}%
    }%
    \gplgaddtomacro\gplbacktext{%
      \put(3315,2041){\makebox(0,0){\strut{} 0}}%
      \put(3523,2041){\makebox(0,0){\strut{} 1}}%
      \put(3730,2041){\makebox(0,0){\strut{} 2}}%
      \put(3937,2041){\makebox(0,0){\strut{} 3}}%
      \put(4144,2041){\makebox(0,0){\strut{} 4}}%
      \put(4352,2041){\makebox(0,0){\strut{} 5}}%
      \put(3833,1711){\makebox(0,0){\strut{}$\Delta M$}}%
    }%
    \gplgaddtomacro\gplfronttext{%
      \csname LTb\endcsname%
      \put(4175,2429){\makebox(0,0)[l]{\strut{}(c)}}%
    }%
    \gplgaddtomacro\gplbacktext{%
      \put(734,482){\makebox(0,0)[r]{\strut{} 0}}%
      \put(734,665){\makebox(0,0)[r]{\strut{} 0.01}}%
      \put(734,847){\makebox(0,0)[r]{\strut{} 0.02}}%
      \put(734,1029){\makebox(0,0)[r]{\strut{} 0.03}}%
      \put(734,1212){\makebox(0,0)[r]{\strut{} 0.04}}%
      \put(933,181){\makebox(0,0){\strut{} 0}}%
      \put(1150,181){\makebox(0,0){\strut{} 0.5}}%
      \put(1367,181){\makebox(0,0){\strut{} 1}}%
      \put(1584,181){\makebox(0,0){\strut{} 1.5}}%
      \put(1801,181){\makebox(0,0){\strut{} 2}}%
      \put(2017,181){\makebox(0,0){\strut{} 2.5}}%
      \put(2234,181){\makebox(0,0){\strut{} 3}}%
      \put(2451,181){\makebox(0,0){\strut{} 3.5}}%
      \put(100,929){\rotatebox{-270}{\makebox(0,0){\strut{}$\Delta E \left( \frac{e^2}{\epsilon \ell_B} \right)$}}}%
      \put(1800,-148){\makebox(0,0){\strut{}$\Delta M$}}%
    }%
    \gplgaddtomacro\gplfronttext{%
      \csname LTb\endcsname%
      \put(2413,561){\makebox(0,0)[l]{\strut{}(d)}}%
    }%
    \gplgaddtomacro\gplbacktext{%
      \put(2676,181){\makebox(0,0){\strut{} 0}}%
      \put(2893,181){\makebox(0,0){\strut{} 0.5}}%
      \put(3110,181){\makebox(0,0){\strut{} 1}}%
      \put(3326,181){\makebox(0,0){\strut{} 1.5}}%
      \put(3543,181){\makebox(0,0){\strut{} 2}}%
      \put(3760,181){\makebox(0,0){\strut{} 2.5}}%
      \put(3976,181){\makebox(0,0){\strut{} 3}}%
      \put(4193,181){\makebox(0,0){\strut{} 3.5}}%
      \put(4410,181){\makebox(0,0){\strut{} 4}}%
      \put(3543,-148){\makebox(0,0){\strut{}$\Delta M$}}%
    }%
    \gplgaddtomacro\gplfronttext{%
      \csname LTb\endcsname%
      \put(4155,561){\makebox(0,0)[l]{\strut{}(e)}}%
    }%
    \gplbacktext
    \put(0,0){\includegraphics{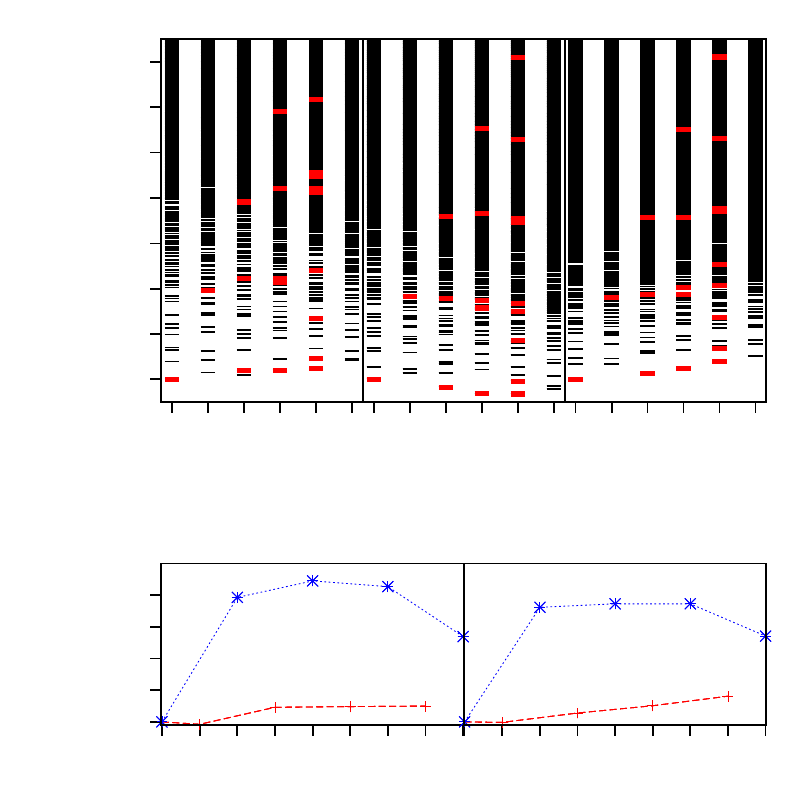}}%
    \gplfronttext
  \end{picture}%
\endgroup
\caption{\label{12ZEdge} (a) The spectrum of 10 particles in 22 states for $d=1.2\ell_B,$ $\kappa=0.3,$ and $w=0\ell_B$ with the edge states highlighted in red.  (b),(c) The same as (a), but for $w=1\ell_B$ at $\kappa=0.4$ and $\kappa=0.5$, respectively.  (d),(e) The energies of the Fermi (red) and Bose (blue) edge modes as a function of angular momentum with $w=0\ell_B$ and $w=1\ell_B$, respectively.}
\end{figure}

We now focus on the Pfaffian and its edge structure, which consists of charged Bose modes and neutral Fermi modes.  This  structure is a signature of a paired ground state as the Fermi edge modes arise by pair breaking excitations and thus their appearance would be a strong indication of the ground state being in the same universality class as the Pfaffian \cite{Edges}.  As our choice of a truncated state space suppresses the edge states and counteracts edge reconstruction, we work in a larger state space. Then we expect edge reconstruction as the system is above the $d=0.5\ell_B$ limit discussed in Ref. \onlinecite{5/2EdgeReconstruction}.  Carrying out the exact diagonalization calculation in this expanded basis, we identify the edge modes by looking first at the overlaps with the Pfaffian edge states and then considering the energies for the states consisting of multiple modes.  The results are shown in Fig. \ref{12ZEdge}a-c.

From Fig. \ref{12ZEdge}a, we have $d=1.2\ell_B,\kappa=0.3$ and $w=0\ell_B.$  In the spectrum, we identify two clear branches of edge modes: an upper and a lower branch.  The lower branch represents the purely Fermi edge modes, which are sufficiently separated from the bulk, though there is still some mixing at $\Delta M=2$.  The upper branch is well mixed with the bulk and consists of the Bose edge modes and the mixed edge modes.

For the finite confinement $w=1\ell_B$ system, we examine $\kappa=0.4$ in Fig. \ref{12ZEdge}b.  At this $\kappa$, the state has undergone edge reconstruction and the Fermi edge modes lie universally below the $M=85$ state.  As we increase LL mixing and look at the $\kappa=0.5$ case, we see that the edge reconstruction is overcome by the effects of LL mixing and $M=85$ is recovered as the ground state.  Thus our choice of a truncated state space has reduced the $\kappa$ at which the incompressible states occur.  Additionally, we see that the Fermi branch is even stronger separated from the bulk states than in the system with $w=0\ell_B.$

With the edge states identified, we calculate the single mode energies.  Comparing the results in Fig. \ref{12ZEdge}d,e for the $w=0\ell_B$ and $w=1\ell_B$, respectively, we see that the Fermi edge mode becomes significantly more linear as finite thickness is introduced.  From this spectrum, we calculate the dispersion relation, which gives the velocities and allows us to calculate the quasiparticle coherence length \cite{CoherenceLength}.  We find $L_\phi\simeq1.93\mu$m and $L_\phi\simeq2.82\mu$m for $w=0\ell_B$ and $w=1\ell_B$ respectively.  These values are lower than previous results \cite{5/2Disk}, though we examine a larger separation $d$.  As larger separations lead to a smoother edge potential, which  lowers the coherence length, we expect to find even smaller coherence lengths experimentally.

In conclusion, our simulations of the 5/2 state on a disk of neutralizing charge with an account of  of Landau Level mixing  and of finite thickness of the quantum well give three primary results. First, we observe a possible phase transition from the anti-Pfaffian to the Pfaffian as the interaction strength $\kappa$ is increased. This dependence on $\kappa$ is the opposite of that obtained in systems with spherical geometry \cite{ED4}, and the difference arises from the inclusion of interactions with the neutralizing disk. At fixed separation $d$, $\kappa$ acts to bend the phases so that they occur at larger $d$ than they originally appear, and the incompressible regions expand as $w$ is increased.  We also find that only the MR and anti-Pfaffian states continue to possess energy gaps at finite well width, while the gaps for the compressible stripe states close. Second, for the charge $e/4$ quasiholes, we found that the quasihole size necessarily increases as a result of the decreasing characteristic magnetic field strength for increasing $\kappa$. Third, the LL mixing is essential for the realization of the Pfaffian state in the expanded state space, as edge reconstruction destroys the signatures of the Pfaffian state for relatively small $d$ otherwise. The edge structure of the Pfaffian is drastically improved when both $\kappa$ and $w$ increase.

With the inclusion of finite thickness, our next aim is exploring what happens when subbands of different LLs become degenerate. Such studies have the potential to shed light on experimental results \cite{TunnelingConduction} which appear to exhibit the Halperin 331 state \cite{Halperin}. Additionally, improvements to this type of calculation can lead to insight into engineering samples which exhibit a desired ground state.

{\it Acknowledgement.} This work was supported by the U.S. Department of Energy, Office of Basic Energy Sciences, Division of Materials Sciences and Engineering under Award
DE-SC0010544.


\begin{thebibliography}{100}

\bibitem{5/2Observation1}
R. Willett, J. P. Eisenstein, H. L. Stormer, D. C. Tsui, A. C. Gossard, and J. H. English, Phys. Rev. Lett. \textbf{59}, 1776 (1987).

\bibitem{5/2Observation2}
W. Pan, R. R. Du, H. L. Stormer, D. C. Tsui, L. N. Pfeiffer, K. W. Baldwin, and K. W. West, Phys. Rev. Lett. \textbf{83}, 820 (1999).

\bibitem{Laughlin}
R. B. Laughlin, Phys. Rev. Lett. \textbf{50}, 1395 (1983).

\bibitem{Jain}
J. K. Jain, Phys. Rev. Lett. \textbf{63}, 199 (1989).

\bibitem{CooperPairing}
V. W. Scarola, K. Park and J. K. Jain, Nature \textbf{406}, 863 (2000).

\bibitem{MooreRead}
G. Moore and N. Read, Nucl. Phys. B \textbf{360}, 362 (1991).

\bibitem{StatPairing1}
M. Greiter, X.-G. Wen, and F. Wilczek, Phys. Rev. Lett. \textbf{66}, 3205 (1991).

\bibitem{StatPairing2}
M. Greiter, X.-G. Wen, and F. Wilczek, Nucl. Phys. B \textbf{374}, 567 (1992).

\bibitem{aPf1}
M. Levin, B. I. Halperin, and B. Rosenow, Phys. Rev. Lett. \textbf{99}, 236806 (2007).

\bibitem{aPf2}
S.-S. Lee, S. Ryu, C. Nayak, and M. P. A. Fisher, Phys. Rev. Lett. \textbf{99}. 236807 (2007).

\bibitem{nonabelian1}
V. Gurarie and C. Nayak, Nucl. Phys. B \textbf{506}, 685 (1997).

\bibitem{nonabelian2}
D. A. Ivanov, Phys. Rev. Lett. \textbf{86}, 268 (2001).

\bibitem{nonabelian3}
Y. Tserkovnyak and S. H. Simon, Phys. Rev. Lett. \textbf{90}, 016802 (2003).

\bibitem{SpontaneousBreaking}
M. R. Peterson, K. Park, and S. Das Sarma, Phys. Rev. Lett. \textbf{101}, 156803 (2008).

\bibitem{Pseudo1}
W. Bishara and C. Nayak, Phys. Rev. B \textbf{80}, 121302 (2009).

\bibitem{Pseudo2}
S. H. Simon and E. H. Rezayi, Phys. Rev. B \textbf{87}, 155426 (2013).

\bibitem{Pseudo3}
M. R. Peterson and C. Nayak, Phys. Rev. B \textbf{87}, 245129 (2013).

\bibitem{Pseudo4}
I. Soderman and A. H. MacDonald, Phys. Rev. B \textbf{87}, 245425 (2013).

\bibitem{Pseudo5}
R. E. Wooten, J. H. Macek, and J. J. Quinn, Phys. Rev. B \textbf{88}, 155421 (2013).

\bibitem{ED1}
A. Wojs, C. Toke, and J. K. Jain, Phys. Rev. Lett. \textbf{105}, 096802 (2010).

\bibitem{ED2}
E. H. Rezayi and S. H. Simon, Phys. Rev. Lett. \textbf{106}, 116801 (2011).

\bibitem{ED3}
Z. Papic, F. D. M. Haldane, and E. H. Rezayi, Phys. Rev. Lett. \textbf{109}, 266806 (2012).

\bibitem{DMRG}
M. R. Zaletel, R. S. K. Mong, F. Pollmann, and E. H. Rezayi, arXiv:1410.3861 (unpublished).

\bibitem{ED4}
K. Pakrouski, M. R. Peterson, Th. Jolicoeur, V. W. Scarola, C. Nayak, and M. Troyer, arXiv:1411.1068 (unpublished).

\bibitem{FiniteThickness1}
M. R. Peterson, Th. Jolicoeur, and S. Das Sarma, Phys. Rev. Lett. \textbf{101}, 016807 (2008).

\bibitem{FiniteThickness2}
M. R. Peterson, Th. Jolicoeur, and S. Das Sarma, Phys. Rev. B \textbf{78}, 155308 (2008).

\bibitem{FiniteThickness3}
Y. Liu, D. Kamburov, M. Shayegan, L. N. Pfeiffer, K. W. West, and K. W. Baldwin, Phys. Rev. Lett. \textbf{101}, 176805 (2011).

\bibitem{1/3Disk}
X. Wan, E. H. Rezayi, and K. Yang, Phys. Rev. B \textbf{68}, 125307 (2003).

\bibitem{5/2Disk}
X. Wan, Z.-X. Hu, E. H. Rezayi, and K. Yang, Phys. Rev. B \textbf{77}, 165316 (2008).

\bibitem{LLMixingHoles}
S. L. Sondhi and S. A. Kivelson, Phys. Rev. B \textbf{46}, 13319 (1992).

\bibitem{5/2EdgeReconstruction}
Y. Zhang, Y.-H. Wu, J. A. Hutasoit, and J. K. Jain, arXiv:1406.7296 (unpublished).
\bibitem{Haldane}
F. D. M. Haldane, Phys. Rev. Lett. \textbf{51}, 605 (1983).

\bibitem{AngularStates}
S. H. Simon, E. H. Rezayi, and N. R. Cooper, Phys. Rev. B \textbf{75}, 195306 (2007).

\bibitem{Transition}
N. Samkharadze, L. N. Pfeiffer, K. W. West, and G. A. Csathy, aRxiv:1302.1444 (unpublished).

\bibitem{Quasiholes}
C. Nayak and F. Wilczek, Nucl. Phys. B \textbf{479}, 529 (1996).

\bibitem{Edges}
M. Milovanovic and N. Read, Phys. Rev. B \textbf{53}, 13559 (1996).

\bibitem{CoherenceLength}
W. Bishara and C. Nayak, Phys. Rev. B \textbf{77}, 165302 (2008).

\bibitem{TunnelingConduction}
X. Lin, C. Dillard, M. A. Kastner, L. N. Pfeiffer, and K. W. West, Phys. Rev. B \textbf{85}, 165321 (2012).

\bibitem{Halperin}
B. I. Halperin, Helv. Phys. Acta. \textbf{56}, 75 (1983).

\end{thebibliography}
\end{document}